\documentclass{IEEEtran}
\usepackage{amsmath}
\usepackage{nccmath}
\usepackage[utf8]{inputenc}
\usepackage{epsfig,scalefnt,multirow}
\usepackage{url}
\usepackage{ifthen}
\usepackage{mathtools}
\usepackage{cite}
\usepackage{graphicx}
\usepackage{amssymb}
\usepackage{tabularx}
\usepackage{amsmath}
\usepackage{epstopdf}
\usepackage{epsf}
\usepackage{algorithm}
\usepackage{algpseudocode}
\usepackage{algpascal}
\usepackage{cases}
\usepackage{subfig}
\usepackage{stfloats}
\usepackage{float}
\usepackage{xcolor}
\usepackage{tabularx}
\usepackage{epsfig,amsmath,amssymb,epsf,amsthm,scalefnt,multirow,subfig}
\usepackage{psfrag}
\usepackage{gensymb}
\usepackage{cleveref}
\usepackage{mathrsfs}
\usepackage{enumerate}
\usepackage[labelsep=period]{caption}

   \def\cP{{\mathcal{P}}}

\def\diag{\mathop{\mathrm{diag}}}

\def\trace{\mathop{\mathrm{tr}}}

\def\b0{{\pmb{0}}} 

 \def\bb{{\mathbf{b}}}  \def\bd{{\mathbf{d}}}
   \def\bh{{\mathbf{h}}}

  \def\bs{{\mathbf{s}}} 
 \def\bv{{\mathbf{v}}} \def\bw{{\mathbf{w}}} \def\bx{{\mathbf{x}}}
\def\by{{\mathbf{y}}} \def\bz{{\mathbf{z}}}  \def\bone{\mathbf{1}}

\def\bA{{\mathbf{A}}}   
  \def\bG{{\mathbf{G}}} \def\bH{{\mathbf{H}}}
\def\bI{{\mathbf{I}}}   
   \def\bP{{\mathbf{P}}}
  \def\bS{{\mathbf{S}}} 
 \def\bV{{\mathbf{V}}} \def\bW{{\mathbf{W}}} \def\bX{{\mathbf{X}}}

\DeclarePairedDelimiter\norm{\lVert}{\rVert}

\begin{document}

\title{ \huge{Cell-Free MIMO Systems Powered by \\ Intelligent Reflecting Surfaces} }

\author{Taegyun~Noh	and~Junil~Choi% 
	\thanks{T. Noh is with Electronics and Telecommunications Research Institute, 
		Daejeon, 34129, South Korea, and also with the School of Electrical Engineering, 
		Korea Advanced Institute of Science and Technology, Daejeon 34141, South Korea 
		(e-mail: taegyun@etri.re.kr).}%
	\thanks{J. Choi is with the School of Electrical Engineering, Korea
		Advanced Institute of Science and Technology, Daejeon 34141, South Korea
		(e-mail: junil@kaist.ac.kr).}% 
}

\maketitle

\begin{abstract} 
Cell-free massive multiple-input multiple-output (MIMO) and intelligent reflecting surface (IRS) are considered as the prospective multiple antenna technologies for beyond the fifth-generation (5G) networks. Cell-free MIMO systems powered by IRSs, combining both technologies, can further improve the performance of cell-free MIMO systems at low cost and energy consumption. Prior works focused on instantaneous performance metrics and relied on alternating optimization algorithms, which impose huge computational complexity and signaling overhead. To address these challenges, we propose a novel two-step algorithm that provides the long-term passive beamformers at the IRSs using statistical channel state information (S-CSI) and short-term active precoders and long-term power allocation at the access points (APs) to maximize the minimum achievable rate. Simulation results verify that the proposed scheme outperforms benchmark schemes and brings a significant performance gain to the cell-free MIMO systems powered by IRSs.
\end{abstract}


\section{Introduction} 
\IEEEPARstart{C}{ell}-free massive multiple-input multiple-output (MIMO) systems have been proposed to effectively alleviate intercell interference by coordinating a large number of distributed access points (APs), which are connected through fronthaul links to the central processing unit (CPU) \cite{NAY+17, NAM+17, ZBM+20}. Intelligent reflecting surface (IRS), also known as reconfigurable intelligent surface (RIS), has been considered as one of the prospective multiple antenna technologies for beyond the fifth-generation (5G) networks \cite{ZBM+20}. By adjusting phase shifts of the IRS elements, the propagation environment can be favorably manipulated.

Cell-free MIMO systems powered by IRSs have been recently introduced to further enhance the performance of cell-free MIMO systems at low and affordable cost and energy consumption by integrating multiple IRSs with the cell-free MIMO systems. The existing works, which studied the active and passive beamforming design, focused on the instantaneous performance metrics, i.e., instantaneous sum-rate \cite{ZD21, ZDZ+21, HYX+21} or energy efficiency \cite{ZDZ+21a, LND20X}. Theses works adopted alternating optimization algorithms, in which the active and passive beamformers are derived based on instantaneous channel state information (I-CSI). Thus, these algorithms would incur huge channel acquisition complexity and related pilot overhead since I-CSI is required for all AP-UE, AP-IRS, and IRS-UE links separately. Moreover, these algorithms would incur immense computational complexity and enormous fronthaul signaling overhead, since the active and passive beamformers need to be computed many times and transferred over fronthaul links for each coherence time. It is also challenging to control the IRSs in real-time, which requires stringent time synchronization \cite{ZBM+20}. These disadvantages can be alleviated by designing a passive beamformer based on long-term channel statistics \cite{HTJ+19, ZWZ+21}.

To the best of our knowledge, this paper is the first attempt for the cell-free MIMO systems powered by IRSs to consider the max-min achievable rate, where the achievable rate is a lower bound of the average rate. This metric aims to provide uniform performance and thus is widely used in the cell-free MIMO systems \cite{NAY+17, NAM+17}. Moreover, we propose a novel non-iterative two-timescale algorithm to obtain 1) short-term active precoders for the APs that depend on I-CSI, 2) long-term power allocation for the APs using statistical CSI (S-CSI), and 3) long-term passive beamformers for the IRSs using S-CSI.

The rest of this paper is organized as follows. In Section~\ref{sec:model}, we present the system model and formulate the max-min achievable rate optimization problem. In Section \ref{sec:two-step}, we propose a two-step algorithm to solve the problem. Simulation results are provided to evaluate the proposed algorithm in Section~\ref{sec:result}. Finally, we conclude the paper in Section \ref{sec:conclusion}.

\textit{Notation:} Vectors and matrices are denoted by lower-case and upper-case boldface letters. The notations $\otimes$ and $\odot$ denote the Kronecker product and Hadamard product. $|\cdot|$ and $\angle(\cdot)$ return the magnitude and angle of a complex argument. $\mathbb{E}\{\cdot\}$ represents the expectation operator. For a square matrix $\bS$, $\trace(\bS)$ denotes the trace operation, and $\bS \succeq 0$ means that $\bS$ is positive semidefinite. A random variable $x \sim \mathcal{CN}(m, \sigma^2)$ is circularly symmetric complex Gaussian (CSCG) distributed with mean~$m$ and variance~$\sigma^2$. 
	
\section{System Model} \label{sec:model}

\begin{figure}[t!] 
	\centering
	\includegraphics[width=0.7 \columnwidth]{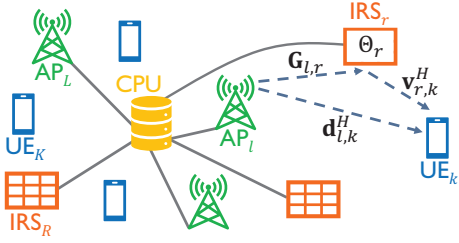}
	\caption{A downlink cell-free MIMO system powered by IRSs.}  \label{fig:model}
\end{figure}

We consider a downlink cell-free MIMO system powered by IRSs as illustrated in Fig. \ref{fig:model}, where $L$ APs and $R$ IRSs are distributed to cooperatively serve $K$ single-antenna user equipments (UEs). All APs and IRSs are connected by wired or wireless fronthaul links to the CPU, which coordinates them. Each AP is equipped with $M$ antennas, and each IRS is comprised of $N$ passive reflecting elements. 

We assume that the data symbols for $K$ UEs $\bs \in \mathbb{C}^{K \times 1}$ are transmitted from all APs \cite{NAY+17, NAM+17}. The transmit signal from AP~$l$ is given by 
\begin{align}
\bx_l = \textstyle\sum_{k=1}^K \bw_{l,k} s_k, 
\end{align}
where $\bw_{l,k} \in \mathbb{C}^{M \times 1}$ and $s_k$ are the active beamforming vector and data symbol for UE $k$. Assuming $\mathbb{E}\{|s_k|^2\} = 1$ for all~$k$, the transmit power constraint of  AP $l$ can be written as
\begin{align}
\textstyle \sum_{k=1}^K \mathbb{E}\{ \norm{\bw_{l,k}}^2 \} \le \bar{P}_l,
\end{align}
where $\bar{P}_l$ denotes the maximum transmit power of AP~$l$.

The channel between an AP and a UE consists of the direct (AP-UE) channel and the $R$ reflection (AP-IRS-UE) channels.\footnote{Note that the signals reflected by the IRSs twice or more are weak enough to be neglected due to the harsh propagation loss of multiple hops \cite{ZD21, ZDZ+21, HYX+21, ZDZ+21a, LND20X}.} Then, the overall channel from AP~$l$ to UE~$k$ can be expressed as
\begin{align}
\bh_{l,k}^H = \bd_{l,k}^H + \textstyle \sum_{r=1}^R \bv_{r,k}^H \boldsymbol{\Theta}_r \bG_{l,r}, \label{eq:overallCh}
\end{align}
where $\bd_{l,k}^H \in \mathbb{C}^{1 \times M}$, $\bG_{l,r} \in \mathbb{C}^{N \times M}$, and $\bv_{r,k}^H \in \mathbb{C}^{1 \times N}$ denote the channel from AP~$l$ to UE~$k$, from AP~$l$ to IRS~$r$, and from IRS~$r$ to UE~$k$, respectively. The reflection coefficient matrix of IRS~$r$ is denoted by $\boldsymbol{\Theta}_r = \diag(\theta_{r,1}, \cdots, \theta_{r,N}) \in \mathbb{C}^{N \times N}$, where  $|\theta_{r,n}| = 1, \forall r,n$ represents the unit-modulus constraint on the IRSs elements.

We assume the Rician fading channel model for all channels \cite{ZD21, ZDZ+21, ZDZ+21a, WZ19}. Specifically, the channel between AP~$l$ and UE~$k$ is given by
\begin{align}
\bd_{l,k} &= \textstyle \sqrt{\xi_{l,k}^\mathrm{d}} \sqrt{\frac{\beta_\mathrm{d}}{1+\beta_\mathrm{d}}}\bar{\bd}_{l,k}' + \sqrt{\xi_{l,k}^\mathrm{d}} \sqrt{\frac{1}{1+\beta_\mathrm{d}}}\tilde{\bd}_{l,k}' \notag \\
&= \bar{\bd}_{l,k} + \tilde{\bd}_{l,k}, \label{eq:ricianCh}
\end{align}
where $\bar{\bd}_{l,k}'$, $\tilde{\bd}_{l,k}'$, and $\beta_\mathrm{d}$ denote the line-of-sight (LoS) component, non-line-of-sight (NLoS) component, and Rician K-factor of the channel~$\bd_{l,k}$, respectively. The channel~$\bd_{l,k}$ includes the distance-dependent path loss $\xi_{l,k}^\mathrm{d}$. The AP-IRS and IRS-UE channels follow the same model as in \eqref{eq:ricianCh} with proper notation changes. It is assumed that the NLoS components of all AP-UE, AP-IRS, and IRS-UE channels are independent each other, and each NLoS component has independent and identically distributed (i.i.d.) $\mathcal{CN}(0, 1)$ entries.

The received signal at UE $k$ can be written as
\begin{align}
y_k &= \textstyle \sum_{l=1}^L \bh_{l,k}^H \bx_l + z_k \notag \\
	&= \bh_{k}^H \bw_{k} s_k + \textstyle \sum_{k' \neq k}^K \bh_{k}^H \bw_{k'} s_{k'} + z_k,
\end{align}
where $\bh_k = [\bh_{1,k}^T, \cdots, \bh_{L,k}^T]^T \in \mathbb{C}^{LM \times 1}$, $\bw_k = [\bw_{1,k}^T, \cdots, \bw_{L,k}^T]^T \in \mathbb{C}^{LM \times 1}$, and $z_k \sim \mathcal{CN}(0, \sigma^2)$ is the i.i.d. complex additive white Gaussian noise (AWGN). To analyze the theoretic performance gain with the IRSs, we assume that the perfect I-CSI of direct and reflection channels is available at the CPU. We also assume that the UE $k$ has the knowledge of the average of effective channel $\mathbb{E}\{ \bh_{k}^H \bw_{k}\}$ and adopt the \emph{hardening bound}, which is widely used in the massive MIMO literature \cite{BS20a}. Then, the achievable rate of UE $k$ is $\log_2(1+\mathsf{SINR}_k)$, where the effective signal-to-interference-plus-noise ratio (SINR) of UE~$k$ is given by
\begin{align}
\mathsf{SINR}_k = \frac{ | \mathbb{E}\{ \bh_{k}^H \bw_{k}\} |^2 }
{ \textstyle \sum_{k'=1}^K \mathbb{E}\{ |\bh_{k}^H \bw_{k'}|^2 \} - | \mathbb{E}\{ \bh_{k}^H \bw_{k}\} |^2 + \sigma^2}. \label{eq:SINR}
\end{align}

In this paper, we aim to maximize the minimum achievable rate by jointly designing active and passive beamformers subject to the per AP transmit power constraint and unit-modulus constraint on the IRSs elements. This optimization problem can be formulated as
\begin{alignat}{3}
& \max_{ \{\bw_{k}\}, \{\boldsymbol{\Theta}_r\} } ~ && \min_{k} \log_2(1+\mathsf{SINR}_k) \label{eq:prbA} \\
&~ \quad~~ \mathrm{s.t.} &&~ C_1: \textstyle \sum_{k=1}^K \mathbb{E}\{ \norm{\bw_{l,k}}^2 \} \le \bar{P}_l, && ~~\forall l, \notag \\
&&&~ C_2: |\theta_{r,n}| = 1, && ~~\forall r, n, \notag
\end{alignat}
where $\{\bw_{k}\}$ and $\{\boldsymbol{\Theta}_r\}$ represent the active and passive beamformers, respectively. The joint optimization of the problem \eqref{eq:prbA} is very challenging since the active and passive beamformers are tightly coupled.

\section{Proposed Two-Step Algorithm} \label{sec:two-step}
In this section, we propose a suboptimal two-step algorithm to solve the problem \eqref{eq:prbA}. We first design an active beamforming technique, which consists of active precoding and power allocation. Then, we design a passive beamforming technique based on S-CSI. Finally, we summarize the proposed algorithm.

\subsection{Active Beamforming Design} 
We decompose an active beamformer into a short-term active precoder and long-term power allocation to reduce computational complexity and fronthaul signaling overhead. We considered a zero-forcing (ZF) precoder since it shows better max-min rate performance than conjugate beamforming precoder in cell-free MIMO systems \cite{NAM+17}. Moreover, the ZF precoder eliminates inter-user interference, and thus it makes the optimal power allocation simple.

We can express the received signal $\by$ for $K$~UEs as
\begin{align}
\by = \bH^H\bW\bs + \bz,
\end{align}
where $\by = [y_1, \cdots, y_K]^T \in \mathbb{C}^{K \times 1}$, $\bH = [\bh_1, \cdots, \bh_K] \in \mathbb{C}^{LM \times K}$, $\bW = [\bw_1, \cdots, \bw_K] \in \mathbb{C}^{LM \times K}$, and $\bz = [z_1, \cdots, z_K]^T \in \mathbb{C}^{K \times 1}$.
The active beamformer can be set as $\bW = \widetilde{\bW} \bP^{\frac{1}{2}}$ with the ZF precoder $\widetilde{\bW} = \bH \left( \bH^H\bH \right)^{-1}$, where the related condition $LM \ge K$ can be easily fulfilled in the cell-free MIMO systems \cite{NAM+17}. The long-term power allocation $\bP = \diag(p_1, \cdots, p_K) \in \mathbb{C}^{K \times K}$ is applied to all APs.

With the ZF precoder and long-term power allocation, the effective SINR \eqref{eq:SINR} is simply reduced to $\frac{p_k}{\sigma^2}$. With a fixed passive beamformer, the problem \eqref{eq:prbA} boils down to the long-term power allocation problem as
\begin{alignat}{3}
& \cP_1: && \max_{ \bP } ~ && \min_{k} \frac{p_k}{\sigma^2} \label{eq:prb1} \\
&&&~~ \mathrm{s.t.} &&~ C_1: \textstyle \sum_{k=1}^K p_k \mathbb{E}\{ \norm{\tilde{\bw}_{l,k}}^2 \} \le \bar{P}_l, ~\forall l, \notag
\end{alignat} 
where $\widetilde{\bW} = [\tilde{\bw}_1, \cdots, \tilde{\bw}_K] \in \mathbb{C}^{LM \times K}$ and $\tilde{\bw}_k = [\tilde{\bw}_{l,k}^T, \cdots, \tilde{\bw}_{L,k}^T]^T \in \mathbb{C}^{LM \times 1}$.

The objective function in the problem \eqref{eq:prb1} forces the power allocation for all UEs to be the same\footnote{Note that the instantaneous transmit power for UE~$k$ at AP~$l$ is equal to $p_k\norm{\tilde{\bw}_{l,k}}^2$, and thus the actual transmit power is different per UE.}, i.e., $p_1 = \cdots = p_K = p^{\mathrm{opt}}$. Under the typical condition that $\bar{P}_1 = \cdots = \bar{P}_L = \bar{P}$ with a fixed $\bar{P}$, the optimal power allocation $p^{\mathrm{opt}}$ is determined by the AP that consumes the largest power for the active precoder, i.e., $\max_l \textstyle \sum_{k=1}^K \mathbb{E}\{ \norm{\tilde{\bw}_{l,k}}^2 \}$. As the largest power for the active precoder reduces, the optimal power allocation increases, and thus the minimum achievable rate improves accordingly.

\subsection{Passive Beamforming Design} 
Based on the proposed active beamforming design, we find that the passive beamformers are irrelevant to the objective function and only related to the transmit power constraint and unit-modulus constraint in the problem (7). Therefore, we can design a long-term passive beamformer to minimize the largest power for the active precoder. The corresponding optimization problem can be formulated as
\begin{alignat}{2}
& \min_{ \boldsymbol{\theta} } ~ && \max_{l} \textstyle \sum_{k=1}^K\mathbb{E}\{ \norm{\tilde{\bw}_{l,k}}^2 \} \\
&~~ \mathrm{s.t.} &&~ C_2: |\theta_{r,n}| = 1, ~\forall r, n, \notag
\end{alignat}
where $\boldsymbol{\theta} = \boldsymbol{\Theta}^H\bone_{RN} \in \mathbb{C}^{RN \times 1}$ and $\boldsymbol{\Theta} = \diag(\boldsymbol{\Theta}_1, \cdots, \boldsymbol{\Theta}_R) \in \mathbb{C}^{RN \times RN}$.

To the best of our knowledge, there is no closed-form expression of $\mathbb{E}\{ \norm{\tilde{\bw}_{l,k}}^2 \}$ in terms of the long-term passive beamformer~$\boldsymbol{\theta}$. It is worth noting, however, that the transmit power reduces as the channel gain increases \cite{WZ19}. Considering this fact, we propose a suboptimal optimization problem that maximizes the minimum average channel gain by passive beamforming at the IRSs as 
\begin{alignat}{2}
& \max_{ \boldsymbol{\theta} } ~ && \min_{k} \textstyle \sum_{l=1}^L\mathbb{E}\{ \norm{\bh_{l,k}}^2 \} \label{eq:prbB} \\
&~~ \mathrm{s.t.} &&~ C_2: |\theta_{r,n}| = 1, ~\forall r, n. \notag
\end{alignat}		

By exploiting S-CSI, the average channel gain of UE~$k$ can be expressed as an explicit function of~$\boldsymbol{\theta}$ that is given as
\begin{align}
\textstyle \sum_{l=1}^L\mathbb{E}\{ \norm{\bh_{l,k}}^2 \}
= \boldsymbol{\theta}^H \bA_k \boldsymbol{\theta} + \boldsymbol{\theta}^H \bb_k + \bb_k^H \boldsymbol{\theta} + c_k, \label{eq:avChGain}
\end{align}
where $\bA_k$, $\bb_k$, and $c_k$ are defined in Appendix A. Since $\bA_k~\succeq~0$, the average channel gain is a convex function of~$\boldsymbol{\theta}$. However, the problem \eqref{eq:prbB} is a non-convex optimization problem since the objective function is not a concave function of~$\boldsymbol{\theta}$, and the unit-modulus constraint is not a convex set.

We apply semidefinite relaxation (SDR) to convert the non-convex problem \eqref{eq:prbB} to a convex problem \cite{NAK+20}. At first, by introducing an auxiliary variable~$q$, the average channel gain of UE~$k$ can be rewritten as 
\begin{align}
\bar{\boldsymbol{\theta}}^H \boldsymbol{\Psi}_k \bar{\boldsymbol{\theta}} + c_k,
\end{align}	
where $\bar{\boldsymbol{\theta}} = \begin{bmatrix} \boldsymbol{\theta} \\ q \end{bmatrix}$ and $\boldsymbol{\Psi}_k = \begin{bmatrix} \bA_k & \bb_k \\ \bb_k^H & 0 \end{bmatrix} \succeq 0$. Note that $\bar{\boldsymbol{\theta}}^H \boldsymbol{\Psi}_k \bar{\boldsymbol{\theta}} = \trace(\boldsymbol{\Psi}_k \bar{\boldsymbol{\theta}} \bar{\boldsymbol{\theta}}^H)$. We define $\bar{\boldsymbol{\Theta}} = \bar{\boldsymbol{\theta}} \bar{\boldsymbol{\theta}}^H$, where $\bar{\boldsymbol{\Theta}} \succeq 0$ and $\mathrm{rank}(\bar{\boldsymbol{\Theta}})=1$. By relaxing the rank-one constraint on $\bar{\boldsymbol{\Theta}}$, which is non-convex, the problem \eqref{eq:prbB} can be reformulated as 
\begin{alignat}{3}
& \cP_2: && \max_{ \bar{\boldsymbol{\Theta}} } ~ && \min_{k} \trace(\boldsymbol{\Psi}_k \bar{\boldsymbol{\Theta}}) + c_k  \label{eq:prb2} \\
&&&~~ \mathrm{s.t.} &&~ [\bar{\boldsymbol{\Theta}}]_{i,i} = 1 , i = 1, \ldots, RN+1, \notag \\
&&&&&~ \bar{\boldsymbol{\Theta}} \succeq 0. \notag
\end{alignat}

Since the problem \eqref{eq:prb2} is a convex semidefinite program (SDP), it can be efficiently solved by existing convex optimization solvers. If the optimal $\bar{\boldsymbol{\Theta}}^{\mathrm{opt}}$ is a rank-one matrix, then the optimal $\bar{\boldsymbol{\theta}}^{\mathrm{opt}}$ is derived by taking the eigenvector corresponding to the maximum eigenvalue of $\bar{\boldsymbol{\Theta}}^{\mathrm{opt}}$. Otherwise, Gaussian randomization is applied to find $\bar{\boldsymbol{\theta}}^{\mathrm{opt}}$ \cite{NAK+20}. Finally, the optimal solution of the problem \eqref{eq:prbB} is recovered by taking ${\boldsymbol{\theta}}^{\mathrm{opt}} = \exp\left( j\angle \left( \left[ \frac{\bar{\boldsymbol{\theta}}^{\mathrm{opt}}}{\bar{\theta}_{RN+1}^{\mathrm{opt}}} \right]_{(1:RN)} \right) \right)$, where $[\bx]_{(1:RN)}$ denotes the vector that contains the first $RN$ entries in $\bx$, and $\bar{\theta}_{RN+1}^{\mathrm{opt}}$ is the last element of~$\bar{\boldsymbol{\theta}}^{\mathrm{opt}}$.

\subsection{Overall Algorithm Description} 
\begin{algorithm}[t]
	\caption{Proposed Two-Step Algorithm}
	\hspace{1mm} \textbf{Input}: S-CSI $\{ \bar{\bd}_{l,k}, \bar{\bG}_l, \bar{\bv}_{k} \}$ \\
	\hspace*{1mm} \textbf{Step 1}: Passive beamforming design
	\begin{itemize} 			
		\item Solve the problem $\cP_2$ to obtain the optimal long-term passive beamformer ${\boldsymbol{\theta}}^{\mathrm{opt}}$.
	\end{itemize}
	\hspace{1mm} \textbf{Step 2}: Active beamforming design
	\begin{itemize} 
		\item Apply the ZF precoder to instantaneous channels with the given ${\boldsymbol{\theta}}^{\mathrm{opt}}$.
		\item Solve the problem $\cP_1$ to obtain the optimal long-term power allocation $\bP^{\mathrm{opt}}$.
	\end{itemize}
\end{algorithm} 
In the previous subsections, we first explained the active beamforming design and then described the passive beamforming design. The proposed algorithm, however, actually operates as summarized in Algorithm 1. First, the optimal long-term passive beamformer ${\boldsymbol{\theta}}^{\mathrm{opt}}$ is obtained by solving $\cP_2$ based on the S-CSI $\{ \bar{\bd}_{l,k}, \bar{\bG}_l, \bar{\bv}_{k} \}$. Then, the ZF precoder is applied to instantaneous channels with the given~${\boldsymbol{\theta}}^{\mathrm{opt}}$, and the optimal long-term power allocation $\bP^{\mathrm{opt}}$ is derived by solving $\cP_1$.

\section{Simulation Results} \label{sec:result}
\begin{figure}[h] 
	\centering
	\includegraphics[width=0.5 \columnwidth]{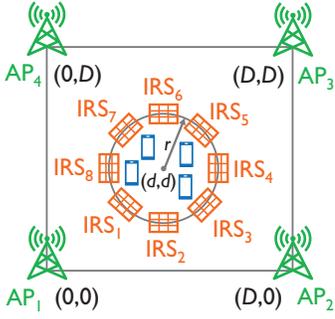}
	\caption{Hotspot deployment scenario.} \label{fig:hotspot}
\end{figure}

In this section, we provide simulation results to validate the minimum achievable rate performance of the proposed two-step algorithm. We consider a hotspot deployment scenario, where the UEs are placed in a hotspot, the APs are deployed a little far from the hotspot, and the IRSs are installed on a circle surrounding the hotspot in order to improve the rate performance \cite{ZD21, ZDZ+21a}. This scenario is illustrated in Fig. \ref{fig:hotspot}, where $L=4$ APs are located at $(0,0)$, $(D,0)$, $(D,D)$, and $(0,D)$, respectively. Up to $R=8$ IRSs are placed on a circle centered at $(d,d)$ with radius $r$, and $K=4$ UEs are uniformly distributed within the circle. We simulate three cases of $d = \{40, 60, 120 \}$~m with $r=30$~m and $D=300$~m.

All APs are equipped with uniform linear arrays (ULAs) at a height of $10$~m with $M = \{4, 8, 16\}$ transmit antennas. Uniform planar array (UPA) is installed at a height of $5$~m for each IRS with $N = \{8, 16, 32, 64, 128\}$ elements. All UEs have single antenna, which is placed at a height of $1.5$~m \cite{M.2412}. We evaluate three cases of $R = \{2, 4, 8\}$, where when $R=2$, only the first and fifth IRSs are present, and when $R=4$, the odd-numbered IRSs are present. It is assumed that all IRSs are deployed on building facades \cite{ZBM+20} and look towards UEs. We consider an AP-IRS blockage model that when a signal from an AP arrives to the back of an IRS, then this signal is not reflected to the UEs, e.g., the signals from the first AP to the first, second, and eighth IRSs are blocked.

The distance-dependent path loss of all channels is modeled as $\xi(d_{\mathrm{link}}) = \xi_0d_{\mathrm{link}}^{-\alpha_\mathrm{X}}$, where $\xi_0$ is the path loss at the reference distance $1$~m, $\alpha_\mathrm{X}$ denotes the path loss exponent of the channel~$\bX$, and $d_{\mathrm{link}}$ represents three-dimensional distance of a channel link considering vertical difference among the APs, IRSs, and UEs. We set $\xi_0 = -30$~dB, $\alpha_\mathrm{d} = 3.4$, $\alpha_\mathrm{v} = \alpha_\mathrm{G} = 2.2$, $\beta_\mathrm{d} = -5$~dB, and $\beta_\mathrm{v} = \beta_\mathrm{G} = 5$~dB considering that the AP-UE channels would suffer from severer attenuation than the AP-IRS and IRS-UE channels \cite{ZWZ+21}. Other system parameters are set as follows: $\bar{P}_l = \{20, 30, 40\}$~dBm for all~$l$, $\sigma^2 = -97$~dBm assuming $10$~MHz of system bandwidth, and $7$~dB of noise figure \cite{M.2412}. We simulate $1,000$ uniform UE drops and generate $1,000$ independent instantaneous channels for each UE drop. 

We consider three benchmark schemes. One is \emph{No-IRS} and another is \emph{Random Passive Beamforming}: the long-term passive beamformers at the IRSs are randomly selected. The active beamforming for both benchmarks is the same as that of the proposed algorithm. To the best of our knowledge, there are no other schemes that can be directly applied to the max-min achievable rate problem in the cell-free MIMO systems powered by IRSs. Instead, we compare with the third benchmark scheme of \emph{Sum-Rate-Max} that maximizes instantaneous sum-rate by utilizing an alternating optimization algorithm to derive the active and passive beamformers [4].

\begin{figure}[t] 
	\centering
	\includegraphics[width=0.95 \columnwidth]{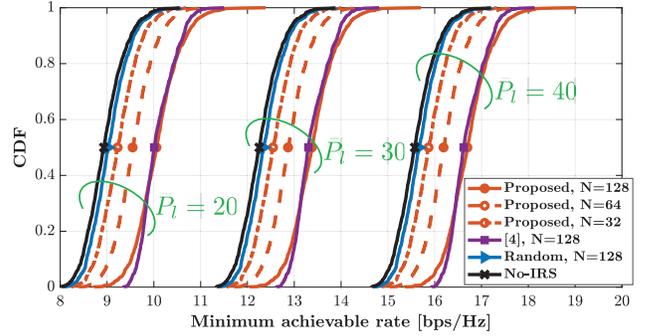}
	\caption{CDFs of the minimum achievable rate by varying the maximum transmit power $\bar{P}_l$ with $d=40$~m, $M=8$, and $R=4$.} \label{fig:cdf}
\end{figure}

Fig. \ref{fig:cdf} depicts the empirical cumulative distribution functions (CDFs) of the minimum achievable rate by varying the maximum transmit power $\bar{P}_l$. The proposed scheme provides a significant gain over No-IRS and the random passive beamforming for all values of $\bar{P}_l$. Specifically for $\bar{P}_l=20$~dBm, the median rate gains of the proposed scheme over No-IRS are equal to $3.4\%$, $7.1\%$, and $12.7\%$ with $N=32$, $64$, and $128$, respectively. By doubling the number of IRS elements, the performance gain is almost doubled. Furthermore, the proposed scheme achieves comparable performance to the Sum-Rate-Max, which requires much higher computational complexity and signaling overhead than the proposed scheme.

\begin{figure*}[t] 
	\centering
	\includegraphics[width=0.9 \textwidth]{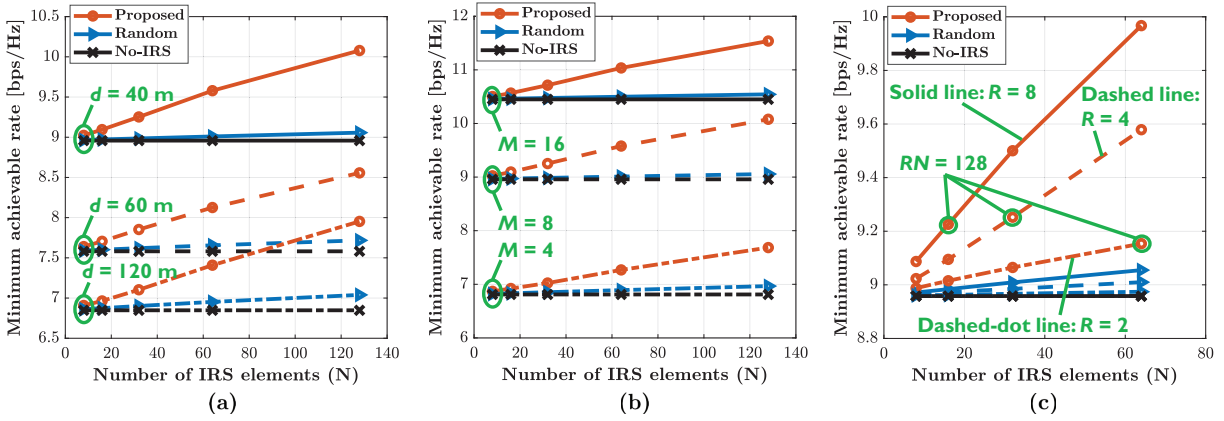}
	\caption{The minimum achievable rate vs. the number of IRS elements $N$: (a) Varying $d$ with $R=4$, $M=8$; (b) Varying $M$ with $d=40$~m, $R=4$; (c) Varying $R$ with $d=40$~m, $M=8$.} 
	\label{fig:rate}
\end{figure*}

In Fig.~\ref{fig:rate}, we plot the minimum achievable rate versus the number of IRS elements~$N$ by varying the center of hotspot~$d$, the number of AP transmit antennas~$M$, and the number of IRSs $R$ with $\bar{P}_l=20$~dBm. For all cases, the minimum achievable rate of the proposed scheme significantly improves with~$N$ and outperforms that of both benchmark schemes.
Fig.~\ref{fig:rate}(a) shows that the performance of all schemes decreases as $d$ increases, i.e., the UEs move towards the center of service area. This is attributed to the fact that the received signal power from the first AP, which is the dominant AP to the UEs, decreases. However, the performance gain of the proposed scheme over No-IRS increases with $d$, i.e., when $N = 128$, the gains are equal to $12.5\%$, $12.9\%$, and $16.1\%$ for $d=40$~m, $60$~m, and $120$~m, respectively.
In Fig.~\ref{fig:rate}(b), it is seen that the smaller $M$, the lower the performance of all schemes.
The proposed scheme, however, provides higher performance gain over No-IRS as $M$ decreases, i.e., when $N = 128$, the gains are equal to $10.4\%$, $12.5\%$, and $12.8\%$ for $M=16$, $8$, and $4$, respectively. 
Fig.~\ref{fig:rate}(c) shows that the performance of the proposed scheme increases with $R$ as expected. 
When the total number of IRS elements $RN$ is the same, similar performance is observed, showing that the proposed scheme is robust to IRS deployment scenarios. 

\section{Conclusion} \label{sec:conclusion}
In this paper, we considered a joint beamforming framework in a cell-free MIMO system powered by IRSs. We formulated a maximization of minimum achievable rate problem and proposed a novel non-iterative two-timescale algorithm that derives the long-term passive beamformers and power allocation and short-term active precoders by exploiting S-CSI. Simulation results revealed that the proposed scheme can significantly improve the minimum achievable rate of the cell-free MIMO systems powered by IRSs compared to the benchmark schemes.

\begin{appendices}
\section{Derivation of \eqref{eq:avChGain}}
The overall channel from AP~$l$ to UE~$k$ in \eqref{eq:overallCh} can be rewritten as 
\begin{align}
\bh_{l,k}^H = \boldsymbol{\theta}^H \bV_k^H \bG_l + \bd_{l,k}^H, 
\end{align}
where, $\bG_l = [\bG_{l,1}^T, \cdots, \bG_{l,R}^T]^T \in \mathbb{C}^{RN \times M}$, $\bV_k^H = \diag(\bv_k^H) \in \mathbb{C}^{RN \times RN}$, and $\bv_k^H = [\bv_{1,k}^H, \cdots, \bv_{R,k}^H] \in \mathbb{C}^{1 \times RN}$. By decomposing the Rician fading channels into the LoS and NLoS components in \eqref{eq:ricianCh}, the average channel gain from AP~$l$ to UE~$k$ can be expressed as
\begin{align}
&\mathbb{E} \{ \norm{\bh_{l,k}}^2 \} \notag \\
&~~= \mathbb{E} \left\{ \left\| (\bar{\bG}_l^H + \tilde{\bG}_l^H)(\bar{\bV}_k + \tilde{\bV}_k) \boldsymbol{\theta} 
	+ (\bar{\bd}_{l,k} + \tilde{\bd}_{l,k}) \right\|^2 \right\} \notag \\
&~~= \boldsymbol{\theta}^H \bA_{l,k} \boldsymbol{\theta} + \boldsymbol{\theta}^H \bb_{l,k} + \bb^H_{l,k}\boldsymbol{\theta} + c_{l,k},
\end{align}
where $\bb_{l,k} = \bar{\bV}_k^H \bar{\bG}_l \bar{\bd}_{l,k}$, $c_{l,k} = \norm{\bar\bd_{l,k}}^2 + \frac{M\xi_{l,k}^\mathrm{d}}{1+\beta_\mathrm{d}}$, and $\bA_{l,k}$ is defined below in \eqref{eq:A_lk}, where $\boldsymbol{\Xi}_k^\mathrm{v} = \diag{(\xi_{1,k}^\mathrm{v}, \cdots, \xi_{R,k}^\mathrm{v}) \in \mathbb{C}^{R \times R}}$, $\boldsymbol{\Xi}_l^\mathrm{G} = \diag{(\xi_{l,1}^\mathrm{G}, \cdots, \xi_{l,R}^\mathrm{G}) \in \mathbb{C}^{R \times R}}$, and $\bA_{l,k} \succeq 0$. All the variables $\bA_{l,k}$, $\bb_{l,k}$, and $c_{l,k}$ are expressed in terms of the S-CSI $\{ \bar{\bd}_{l,k}, \bar{\bG}_l, \bar{\bv}_{k} \}$ and path loss $\{ \xi_{l,k}^\mathrm{d}, \boldsymbol{\Xi}_l^\mathrm{G}, \boldsymbol{\Xi}_k^\mathrm{v} \}$ of all channel links \cite{HTJ+19, ZWZ+21}. The details are omitted here due to the space limitation. Finally, the average channel gain of UE~$k$ can be written as	
\begin{align}
\textstyle \sum_{l=1}^L\mathbb{E}\{ \norm{\bh_{l,k}}^2 \}
= \boldsymbol{\theta}^H \bA_k \boldsymbol{\theta} + \boldsymbol{\theta}^H \bb_k + \bb^H_k \boldsymbol{\theta} + c_k,
\end{align}	
where $\bA_k = \textstyle \sum_{l=1}^L \bA_{l,k}$, $\bb_k = \textstyle \sum_{l=1}^L \bb_{l,k}$, and $c_k = \textstyle \sum_{l=1}^L c_{l,k}$.

\begin{figure*} [b]	
	\hrule \vspace{5mm}	
	\begin{align}
		\bA_{l,k} = \bar{\bV}_k^H \left( \bar{\bG}_l \bar{\bG}_l^H + \frac{M}{1+\beta_\mathrm{G}} \boldsymbol{\Xi}_l^\mathrm{G} \otimes \bI_N \right) \bar{\bV}_k
		+ \left( \frac{1}{1+\beta_\mathrm{v}} \boldsymbol{\Xi}_k^\mathrm{v} \otimes \bI_N \right) \odot \left( \bar{\bG}_l \bar{\bG}_l^H \right) 
		+ \frac{1}{1+\beta_\mathrm{v}}\frac{M}{1+\beta_\mathrm{G}} \boldsymbol{\Xi}_k^\mathrm{v} \boldsymbol{\Xi}_l^\mathrm{G} \otimes \bI_N \label{eq:A_lk}
	\end{align}
\end{figure*}

\end{appendices}

\end{document}